\documentclass[twocolumn,aps]{revtex4}
\usepackage{amsmath}
\usepackage{amssymb}
\usepackage{graphicx}
\usepackage{dcolumn}
\usepackage{bm}
\usepackage{xcolor}
\usepackage{epstopdf}
\usepackage{longtable}
\usepackage{float}
\begin{document}

\preprint{Phys.Rev.B }

\title{Corbino-Enhanced Supersonic Acoustic-Emission Threshold in a  GaAs Two-Dimensional Electron System}
\author{A. D. Levin,$^1$   A. S. Jaroshevich,$^{2}$  Z. D. Kvon,$^{2,3}$  V. A. Chitta,$^1$ M. S. Aksenov, $^{2}$  D. V. Dmitriev,$^{2}$ A. K. Bakarov$^{2}$ and G. M. Gusev$^1$}

\affiliation{$^1$Instituto de F\'{\i}sica da Universidade de S\~ao
Paulo, 135960-170, S\~ao Paulo, SP, Brazil}
\affiliation{$^2$Institute of Semiconductor Physics, Novosibirsk
630090, Russia}
\affiliation{$^3$Novosibirsk State University, Novosibirsk 630090,
Russia}

\date{\today}
\begin{abstract}

We report nonlinear differential-resistance measurements in a high-mobility GaAs/AlGaAs two-dimensional electron system patterned in a Corbino geometry. At zero magnetic field, $R_{\mathrm{diff}}(I_{dc})$ exhibits a pronounced polarity-selective threshold peak on the negative-current branch, gradually suppressed by a perpendicular magnetic field. 
We interpret the Corbino anomaly as a local Cherenkov-like threshold for acoustic phonon emission, enabled by the radial current concentration $j_r(r)=I/(2\pi r)$, which drives the local electron velocity above the sound velocity in a region adjacent to the inner contact. The polarity selectivity is attributed to Peltier heating and cooling, which modify the local thermal and boundary conditions near the contact, rather than to the kinematic threshold itself.
Our results identify the Corbino geometry as a sensitive platform for probing local nonequilibrium electron--phonon processes in a high-mobility electron system within an independently established hydrodynamic-crossover regime and suggest that the inner Corbino contact can act as a geometry-defined source for supersonic acoustic emission.

\end{abstract}

\maketitle

\section{Introduction}

Electron transport in high-mobility two-dimensional electron systems is usually described in terms of momentum relaxation by disorder, phonons, and electron-electron collisions. In sufficiently clean samples, however, electron-electron scattering can become the fastest relaxation process over an extended temperature range. In this case the electronic system reaches local equilibrium before momentum is lost to the lattice, and transport is more naturally described as the flow of a viscous electron fluid with a local velocity field \(\mathbf{u}(\mathbf{r})\)~\cite{narozhny,hui,gurzhi,andreev}. Under large dc bias, this flow can be driven far from equilibrium. When the local electron-fluid velocity becomes comparable to the phase velocity of acoustic excitations, electron-phonon energy relaxation can acquire a threshold-like character analogous to Cherenkov radiation~\cite{kaganov,hutson,wang}.

Supersonic acoustic emission by drifting carriers has recently attracted renewed interest in high-mobility low-dimensional systems~\cite{komirenko,shinokita}. In high-mobility graphene channels, electrically driven carriers can amplify acoustic phonons once the drift velocity exceeds the sound velocity, producing a strong direction-dependent growth of the local resistivity along the carrier-flow direction~\cite{aaron}. These experiments provide a direct example of acoustic phonon amplification producing large nonlinear transport effects in a two-dimensional conductor.

In GaAs-based two-dimensional systems, the dc Hall field, Landau quantization, and acoustic phonon emission combine to produce phonon-induced resistance oscillations and resonant structures in the differential resistivity~\cite{zudov}. Recent experiments on ultrahigh-mobility GaAs/AlGaAs Hall bars demonstrated pronounced magnetophonon features in the supersonic regime, with the sound-barrier condition \(v_{\mathrm{drift}}\simeq s\) reached at a well-defined current density, where \(v_{\mathrm{drift}}\) and \(s\) are the drift and sound velocities, respectively~\cite{studenikin}.

Similar ideas have also been explored in engineered GaAs/AlAs acoustic cavities, where high-\(Q\) phonon confinement can enhance Cherenkov-type phonon amplification~\cite{sujakala}. Together, these results establish supersonic electronic motion as an efficient route for generating and amplifying acoustic phonons in low-dimensional systems~\cite{andersen,eaves,hu}.

Most experiments on supersonic acoustic emission in two-dimensional electron systems have been performed in Hall-bar-like geometries. In such devices the current density is approximately uniform, \(j\simeq I/W\), and the relevant drift velocity is \(v_{\mathrm{drift}}^{\mathrm{HB}}=I/(neW)\), where \(n\) is the carrier density and \(W\) is the channel width. Hall bars are therefore well suited for studying spatially extended, nearly homogeneous nonequilibrium transport or downstream phonon amplification along a channel. They are less sensitive, however, to boundary-localized nonequilibrium regions, because the source and drain contacts are approximately equivalent and the current density is not strongly concentrated at a particular contact.

The Corbino geometry provides a qualitatively different situation~\cite{levchenko,tomadin,afrose,gervais,gervais2}. In a Corbino disk the current flows radially, and the current density is inherently nonuniform,
\[
j_r(r)=\frac{I}{2\pi r}.
\]
Within a hydrodynamic description, the corresponding radial electron-fluid velocity is
\[
u_r(r)=\frac{I}{2\pi rne}.
\]
Thus the largest current density and the largest local flow velocity occur near the inner contact. The inner contact is therefore the natural region where a local supersonic condition can be reached first. In addition, the inner and outer contacts are not geometrically equivalent. 
This geometric inequivalence makes contact-region energy transport and boundary conditions particularly important. Hydrodynamic boundary theories show that electrochemical-potential and temperature drops may become localized near current-penetrable contacts, even when the bulk radial flow remains smooth~\cite{falkovich,kumar2,levchenko2,raichev}. Such boundary localization can influence the visibility and polarity dependence of a nonlinear threshold without making the bare kinematic condition \(\lvert u\rvert=s\) itself current-direction dependent.

In this work we report nonlinear differential-resistance measurements in a high-mobility GaAs/AlGaAs two-dimensional electron system patterned in a Corbino geometry. 
The measurements are performed in a material system in which viscous electron transport has been established by previous magnetotransport studies. Linear-response magnetoresistance measurements on the present Corbino devices also indicate a crossover between viscous and drift-dominated transport; a detailed analysis of this crossover will be reported separately \cite{levin2026}. For the present Corbino devices, the extracted Gurzhi length decreases
from approximately $2$--$3~\mu\mathrm{m}$ at low temperature to
$0.6$--$0.8~\mu\mathrm{m}$ near $60~\mathrm{K}$, while the
electron--electron contribution to the second-moment relaxation rate
follows an approximately $T^2$ dependence~\cite{levin2026}.
These values characterize a viscous-to-drift crossover rather than an
ideal hydrodynamic limit.
The hydrodynamic context of the present experiment is supported by our previous magnetotransport studies of the same GaAs material system~\cite{gusev1, gusev2, levin1, levin2}.
We observe a pronounced current-asymmetric peak in
\[
R_{\mathrm{diff}}=\frac{dV}{dI},
\]
appearing at a threshold negative dc bias and suppressed by perpendicular magnetic field.
We associate the threshold-like contribution with the Corbino-enhanced local velocity near the inner contact. The polarity selectivity is not attributed to the convective term or to the condition \(\lvert u\rvert=s\), both of which are even under current reversal. Instead, it is attributed to Peltier heating and cooling at the contacts, which modify the local thermal and boundary conditions and determine the polarity for which the threshold contribution becomes most visible.

\begin{figure}
 \includegraphics[width=8cm]{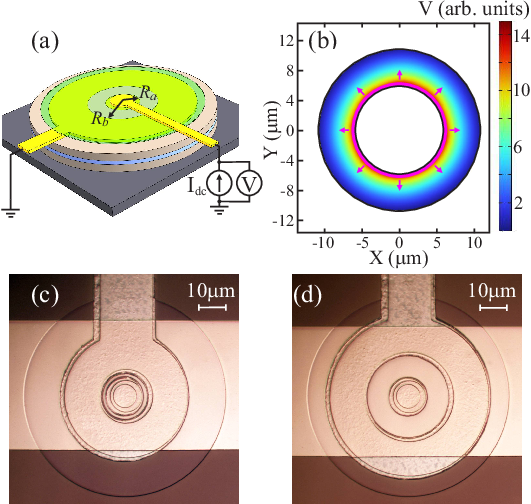}
\caption{(Color online) (a) Schematic of the Corbino disk geometry. 
(b) Calculated radial-flow profile in a Corbino device; acoustic-phonon emission near the inner contact is indicated by red arrows.}
(c) Optical image of the small-ring sample. 
(d) Optical image of the large-ring sample.
\end{figure}

Our results identify the Corbino geometry as a sensitive platform for detecting local nonequilibrium electron--phonon processes in a high-mobility two-dimensional electron system. Hydrodynamics is treated here as the independently established transport context and as a framework emphasizing the special role of the current-penetrable contact region, rather than as a necessary condition for Cherenkov-like acoustic emission.
More broadly, the results suggest that Corbino devices may serve as contact-defined acoustic phonon emitters, where the inner contact acts as a localized source region for supersonic acoustic emission. This geometry may provide a useful starting point for future devices that combine high-mobility electron systems with lithographically defined or epitaxially grown acoustic resonators.

\section{EXPERIMENTAL RESULTS}
\begin{figure*} 
 \includegraphics[width=18cm]{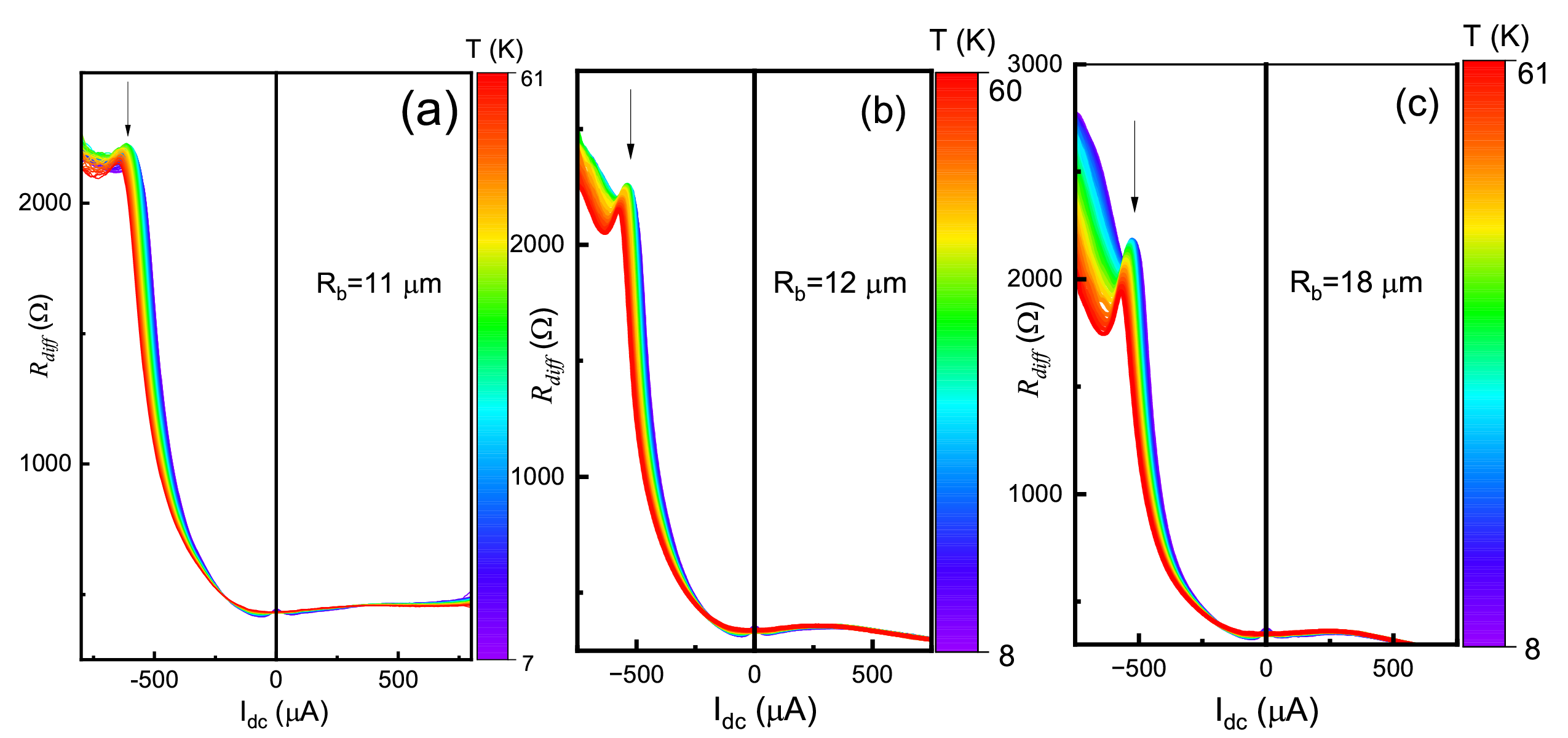}
\caption{(Color online)  Temperature evolution of the differential resistance
$R_{\mathrm{diff}}$ as a function of dc bias current for Corbino devices
with $R_b=11~\mu\mathrm{m}$ (a), $R_b=12~\mu\mathrm{m}$ (b), and
$R_b=18~\mu\mathrm{m}$ (c).Each curve corresponds to a different bath temperature. Panels (a), (b), and (c) contain 27, 26, and 27 traces, respectively,
measured in steps of $2~\mathrm{K}$. Arrows indicate the positions of the peaks on the negative-current branch.}
\end{figure*}

We used high-quality GaAs quantum wells to fabricate the devices studied in this work. The quantum wells had a width of \(14~\mathrm{nm}\), an electron density of approximately
\(n_s=7.1\times10^{11}~\mathrm{cm^{-2}}\) at \(T=4.2~\mathrm{K}\), and a low-temperature mobility of
\(\mu=2\times10^{6}~\mathrm{cm^2/Vs}\).

The main measurements were performed on Corbino devices. Each device consisted of concentric inner and outer ohmic contacts defining an annular current path in the two-dimensional electron system. The inner radius was fixed at \(R_a=6~\mu\mathrm{m}\) for all devices, whereas the outer radius was varied between \(R_b=11\), \(12\), and \(18~\mu\mathrm{m}\). Optical images and schematics of the Corbino disks are shown in Fig.~1.

The Corbino annulus was not defined by mesa etching. Instead, concentric Ti/Ni/Au ohmic contacts were deposited directly on the GaAs/AlGaAs heterostructure and subsequently annealed to establish electrical contact with the two-dimensional electron system. A \(140\)-nm-thick \(\mathrm{SiO_2}\) insulating layer was deposited beneath the metal lead connecting the central contact. The lead was fabricated on top of this dielectric layer to prevent electrical contact with the underlying two-dimensional electron system outside the central electrode and thereby avoid a short circuit between the inner and outer contacts.

\begin{figure*}
    \includegraphics[width=18cm]{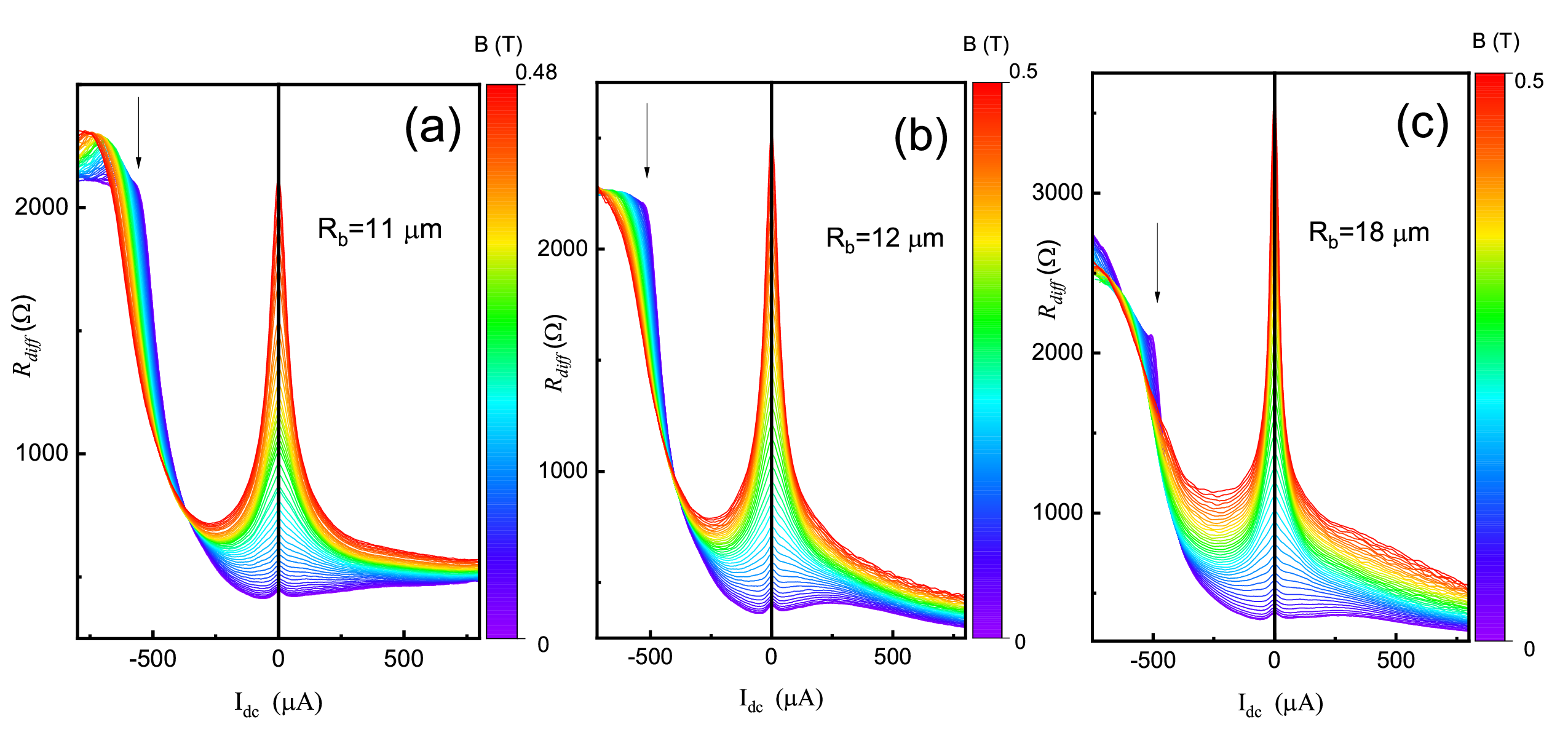}
    \caption{(Color online) Magnetic-field evolution of the differential resistance
    \(R_{\mathrm{diff}}\) as a function of dc bias current for Corbino devices
    with \(R_b=11~\mu\mathrm{m}\) (a), \(R_b=12~\mu\mathrm{m}\) (b), and
    \(R_b=18~\mu\mathrm{m}\) (c). All curves were measured at a bath temperature of \(T= 7~\mathrm{K}\).  Each curve corresponds to a different perpendicular magnetic field,  with a field step of \(\Delta B=0.01~\mathrm{T}\). Panels contain 49 (a) and 51 (b,c) traces.
    Arrows indicate the positions of the peaks on the negative-current branch.}
\end{figure*}

The contact resistance of the Corbino devices was relatively large, typically \(300\)--\(400~\Omega\), and considerably exceeded the intrinsic resistance of the two-dimensional electron system in the annulus. The sheet resistivity of the two-dimensional electron system is \(\rho_0=1/(n_se\mu)\simeq4~\Omega\). The corresponding intrinsic Corbino resistance of the annulus is \(R_{\mathrm{2DES}}=(\rho_0/2\pi)\ln(R_b/R_a)\), giving \(R_{\mathrm{2DES}}\simeq0.4\)--\(1~\Omega\) for our geometries. This large contact resistance is an inherent limitation of the two-terminal Corbino geometry compared with multiterminal Hall-bar measurements, where the contact contribution is excluded from the measured longitudinal resistance. Therefore, a contact-related contribution to the measured resistance cannot be ruled out a priori. Nevertheless, the current-voltage characteristics remained linear at both zero and finite magnetic field, confirming ohmic operation of the contacts over the measurement range. As discussed below, the main experimental trends are not consistent with a predominantly contact-driven interpretation.
For comparison, we also performed measurements on narrow-channel Hall-bar devices fabricated from the same wafer. The Hall-bar geometry was optimized for multiterminal measurements and consisted of three consecutive channel segments with lengths of \(6~\mu\mathrm{m}\), \(20~\mu\mathrm{m}\), and \(6~\mu\mathrm{m}\), each having a width of \(6~\mu\mathrm{m}\). The device incorporated ten voltage probes.

Transport measurements were carried out in a VTI cryostat using direct voltage measurements. For the linear-response measurements, a dc current of \(1~\mu\mathrm{A}\) was applied through the sample, which was sufficiently small to avoid significant electron overheating. For nonlinear measurements, a low-frequency ac excitation in the range
$0.1$--$1~\mu\mathrm{A}$ was applied to measure the differential response. 

At the maximum ac excitation of $1~\mu\mathrm{A}$, the total ac power
dissipated in the two-terminal device is below approximately
$0.4~\mathrm{nW}$, while the power dissipated in the intrinsic Corbino
annulus is below approximately $1~\mathrm{pW}$. Using the low-current
electron-temperature calibration and the approximately quadratic
current-even heating response, we estimate the corresponding increase
in electron temperature to be below $10^{-3}~\mathrm{K}$. The ac
excitation therefore produces negligible heating on the temperature
scale of the present measurements.
The differential resistance was defined as
$R_{\mathrm{diff}}=dV/dI$, where $V$ is the measured voltage response
to the ac excitation. In parallel, Hall-bar devices fabricated from the same wafer, including those used in our previous work~\cite{gusev1, gusev2, levin1,levin2}, were measured to independently determine the parameters of the two-dimensional electron system at zero and finite magnetic field.

Throughout the manuscript, the sign of $I_{\mathrm{dc}}$ is defined with respect to the physical direction of conventional current: positive current flows from the inner contact to the outer contact. This convention is retained when comparing the original and reversed wiring configurations.

Although the contact resistance in the Corbino devices was almost two orders of magnitude larger than the intrinsic resistance of the two-dimensional electron system, the magnetoresistance signal was large and could be extracted unambiguously from the measured magnetic-field dependence.

Figure~2 shows the differential resistance of three Corbino devices with different outer radii as a function of dc current for both current polarities and several temperatures. 

Measurements were also performed on two additional samples with outer radii
$R_b=12~\mu\mathrm{m}$ and $R_b=18~\mu\mathrm{m}$, as shown in
Fig.~S1 of the Supplemental Material~\cite{suppl}. Their
temperature-dependent nonlinear response is very similar to that observed
in the corresponding devices shown in Fig.~2..  All devices exhibit a pronounced nonlinear response with a strong asymmetry with respect to the direction of the dc current. Over a broad current range, the differential resistance measured for negative current polarity substantially exceeds that measured for positive polarity, $R_{\mathrm{diff}}(-I) \gg R_{\mathrm{diff}}(+I)$. The positive-current branch shows only a weak temperature dependence, whereas the negative-current branch changes strongly with temperature. This indicates that the dominant nonlinear contribution is associated with a polarity-selective process that is activated for negative dc bias.

To verify that the polarity-selective anomaly is associated with the physical direction of current flow rather than with the instrumental polarity or wiring configuration, we performed an additional control measurement on the $R_b=18~\mu\mathrm{m}$ Corbino device. Both the current and voltage connections were physically interchanged between the inner and outer contacts. As shown in Fig.~S2 of the Supplemental Material~\cite{suppl}, the
anomaly reverses with respect to the instrumental current convention.
After accounting for the reversed contact configuration, the measurements
coincide within experimental accuracy when expressed using the same
physical current direction relative to the inner contact.
To quantify this polarity-dependent nonlinear response, we introduce a differential diodicity parameter by analogy with fluidic diodes and Tesla valves, where diodicity is commonly defined as the ratio of reverse to forward flow resistance. In the present electronic system we define $\mathcal{D}(I,B,T)
=\frac{R_{\mathrm{diff}}(-|I|,B,T)}{R_{\mathrm{diff}}(+|I|,B,T)}$. With this convention, \(\mathcal{D}=1\) corresponds to a current-symmetric response, while \(\mathcal{D}>1\) indicates a larger differential resistance for the negative-current polarity. Since the sign convention for current is arbitrary, the numerator is chosen to correspond to the polarity for which the Corbino anomaly is observed.

For all Corbino geometries, we estimate a large diodicity, $\mathcal{D}\approx 6$,
near \(|I_{\mathrm{dc}}|\simeq 500~\mu\mathrm{A}\) at a bath temperature of \(T=8~\mathrm{K}\). 

\begin{figure} 
 \includegraphics[width=8cm]{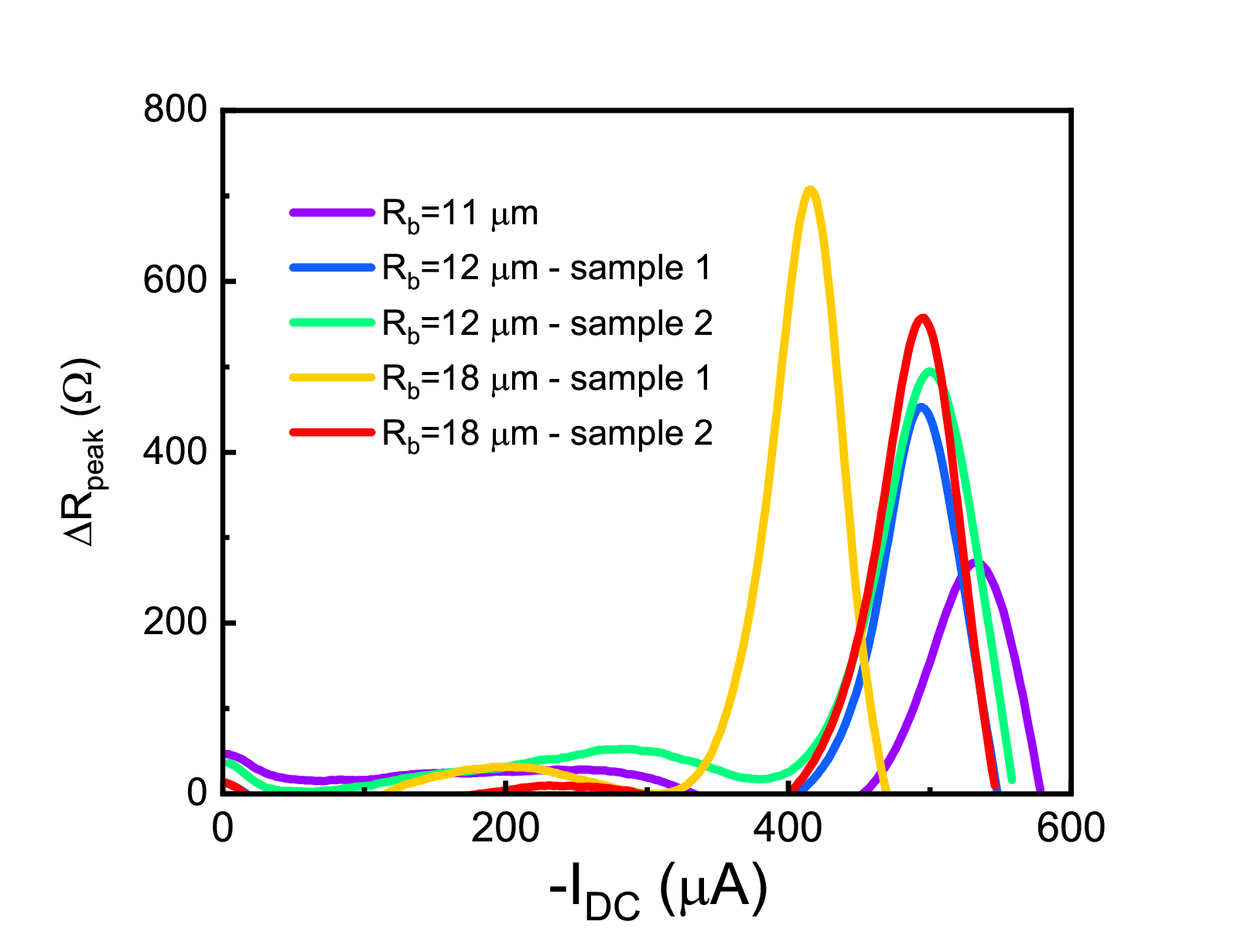}
\caption{
(Color online) Background-subtracted threshold-like contribution
$\Delta R_{\mathrm{peak}}=R_{\mathrm{diff}}-R_{\mathrm{bg}}$
as a function of $-I_{\mathrm{dc}}$ for the negative-current branch
of Corbino devices with different outer radii at $T=7~\mathrm{K}$.
The background $R_{\mathrm{bg}}$ was obtained using the same
third-order polynomial procedure described in the text.
Curves marked ``sample 1'' and ``sample 2'' correspond to independent
nominally identical devices. The corresponding temperature-dependent
measurements for the devices labeled ``sample 2'' are shown in
Fig.~S1 of the Supplemental Material~\cite{suppl}.
}
\end{figure}

\begin{figure}
    \includegraphics[width=8cm]{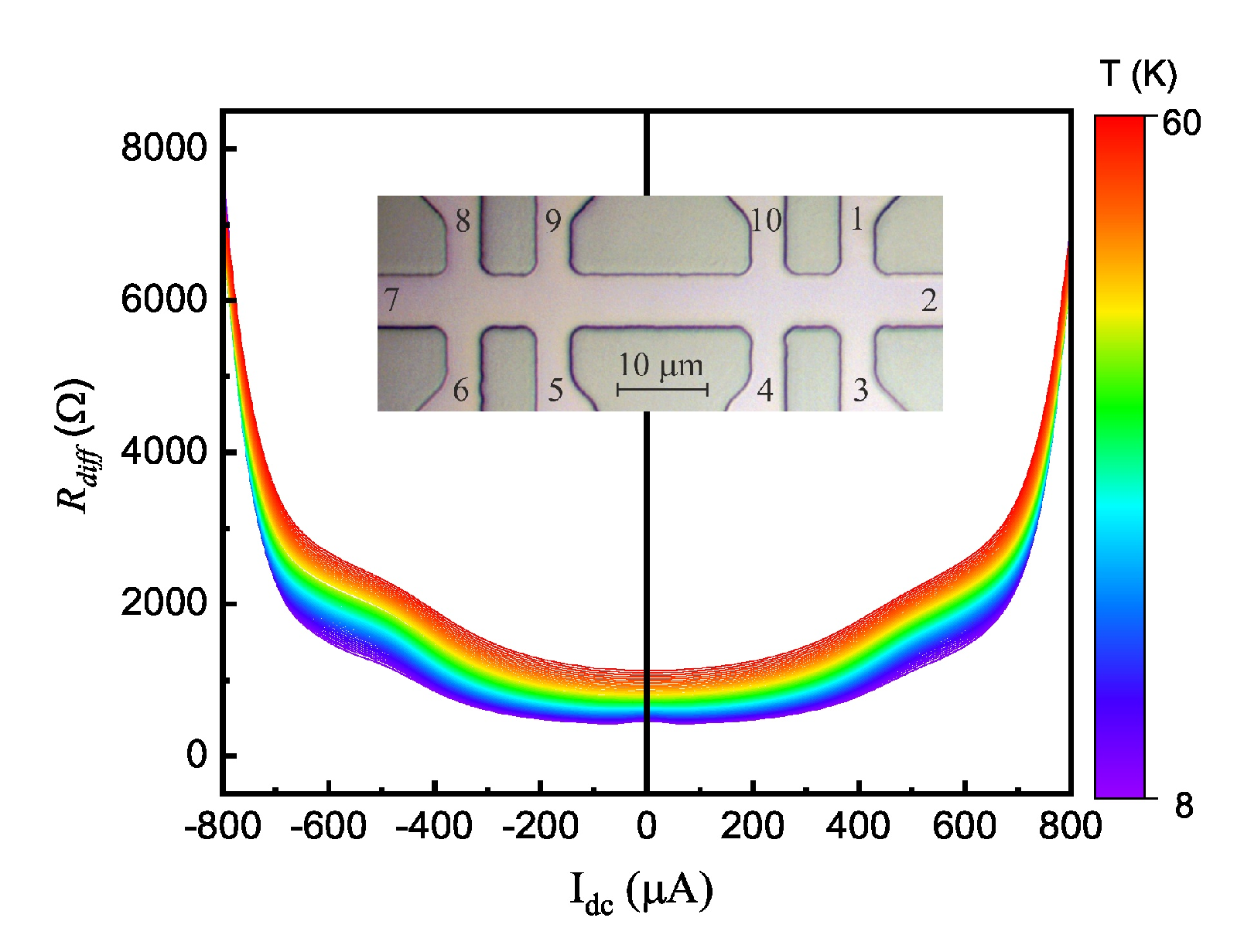}
    \caption{
    (Color online) Differential resistance
    \(R_{\mathrm{diff}}=dV/dI\) of the Hall-bar control device as a function
    of dc bias current \(I_{\mathrm{dc}}\). The panel contains 27 traces measured
from $T=8$ to $60~\mathrm{K}$ in steps of $2~\mathrm{K}$.
The inset shows a schematic of the Hall-bar geometry and the measurement
    configuration. Current was applied between contacts 7 and 2, while the differential
voltage was measured between contacts 5 and 4.
 The Hall-bar response is broad and nearly symmetric under current reversal,
    with no sharp polarity-selective feature comparable to that observed in
    the Corbino devices.}
\end{figure}
 The pronounced threshold-like anomaly observed on the negative-current branch at zero magnetic field is attributed to the onset of local acoustic phonon emission near the inner Corbino contact. In this region the radial current density is maximal, and the local electron flow can approach the acoustic velocity. We therefore associate the anomaly with a Cherenkov-like emission process enhanced by the Corbino geometry. A detailed physical interpretation, including the role of hydrodynamic flow, the localized electrochemical-potential drop, and the phenomenological threshold model, is presented below in the Discussion section.

Figure~3 shows the evolution of the nonlinear differential resistance of the three Corbino devices with perpendicular magnetic field. The most prominent field-induced effect is the development of a strong enhancement of \(R_{\mathrm{diff}}\) near zero dc current. This near-zero-current peak is consistent with the large positive magnetoresistance of the Corbino devices, which is approximately quadratic in magnetic field in the low-field range. As \(B\) is increased, the zero-bias resistance grows rapidly and becomes the dominant contribution to the low-current part of the nonlinear response.
At the same time, the anomalous threshold peak observed on the
negative-current branch is progressively suppressed by magnetic field.

The narrow threshold-like peak is progressively suppressed and broadened,
whereas the broader high-current asymmetry remains visible.
To test the magnetic-field-reversal symmetry directly, we performed additional measurements on the representative $R_b=18~\mu\mathrm{m}$ Corbino device at equal positive and negative perpendicular magnetic fields. As shown in Fig.~S3 of the Supplemental Material~\cite{suppl}, the
$R_{\mathrm{diff}}(I_{\mathrm{dc}})$ curves, including the field-induced
suppression of the threshold-like peak, are symmetric within experimental
accuracy under $B\rightarrow-B$.

This behavior supports the interpretation that the threshold anomaly is a local Corbino-specific effect rather than a simple bulk nonlinearity. A perpendicular magnetic field modifies the electron trajectories and the local electrochemical-potential distribution near the contacts. As a result, the localized threshold contribution associated with the inner Corbino contact is suppressed, while the conventional positive Corbino magnetoresistance produces the strong near-zero-current enhancement.

Figure~4 shows the background-subtracted threshold-like contribution,
\[
\Delta R_{\mathrm{peak}}(I)
=
R_{\mathrm{diff}}(I)-R_{\mathrm{bg}}(I),
\]
for Corbino devices with different outer radii.

To extract this contribution, the full negative-current branch over the
interval $-800\leq I_{\mathrm{dc}}\leq0~\mu\mathrm{A}$ was fitted with a
third-order polynomial, which was used as a smooth phenomenological
background for all datasets.

A clear residual peak is observed for every Corbino device. Its position is
sample dependent and lies within
$|I_{\mathrm{dc}}|\simeq400$--$600~\mu\mathrm{A}$. Importantly, the peak position does not show a simple monotonic dependence on the outer radius of the Corbino ring. This indicates that the threshold is not determined solely by the global annular width or by the outer contact position. Instead, it is likely controlled by the local nonequilibrium conditions near the inner Corbino contact, including current crowding, local heating, boundary relaxation, and the effective electrochemical-potential drop.

The amplitude of the extracted peak also varies between devices, but no universal scaling with the Corbino diameter is observed. Thus, while the broad current asymmetry is a robust feature of all Corbino geometries, the narrow threshold peak is sensitive to sample-specific boundary conditions. This behavior is consistent with a local contact-assisted process rather than a homogeneous bulk instability. In this interpretation, the peak reflects the onset of an additional dissipation channel localized near the inner contact, while its precise position and magnitude depend on the microscopic state of the boundary region.

Figure~5 shows the differential resistance of the narrow Hall-bar device as a function of dc current for several temperatures. In this configuration, the current \(I\) was driven between contacts 7 and 2, while the voltage \(V\) was measured between probes 5 and 4, yielding $R_{\mathrm{diff}}=\frac{dV_{5,4}}{dI_{7,2}}$, as shown in the inset of Fig.~5. In contrast to the Corbino data, the Hall-bar response is nearly symmetric with respect to current reversal. The differential resistance increases strongly at large \(|I_{\mathrm{dc}}|\), forming broad nonlinear features on both current branches. These structures become more pronounced with increasing temperature, but they remain smooth and strongly broadened rather than developing into a sharp threshold peak. The overall shape is therefore consistent with a generic nonlinear response of the high-mobility GaAs/AlGaAs two-dimensional electron system at large dc bias, possibly involving current-induced electron heating, nonlinear electron-phonon relaxation, or a weak contribution from acoustic phonon emission. However, because the features are broad and nearly symmetric in current, they do not provide a unique signature of a local supersonic emission threshold.

The Hall-bar measurement therefore serves as an important control experiment. It demonstrates that nonlinear differential resistance can occur in the same material system without producing the sharp, polarity-selective anomaly observed in the Corbino geometry. The Corbino response is qualitatively different: the anomaly is narrow, strongly asymmetric, and associated with a specific current polarity. This comparison supports the interpretation that the Corbino peak is not a generic bulk nonlinearity of the two-dimensional electron system, but rather a geometry-enhanced local threshold associated with the inner Corbino contact.

\section{Discussion}

The nonlinear Corbino response is characterized by three key observations: a narrow, polarity-selective peak in $R_{\mathrm{diff}}$ on the negative-current branch; suppression and smoothing of this peak by a perpendicular magnetic field; and the absence of a comparable sharp polarity-selective feature in a Hall-bar device fabricated from the same wafer. In this section we develop a microscopic interpretation of these observations and provide a quantitative estimate of the visibility factor $\alpha_C$ that relates the observed peak current $I_{\mathrm{peak}}$ to the bare kinematic threshold $I_{u=s}$.

\subsection{Physical picture: boundary-assisted Cherenkov-like emission at the inner contact}

We interpret the Corbino anomaly as a local, boundary-assisted threshold for acoustic phonon emission.

The kinematic onset of emission and the observed polarity selectivity are treated as two physically distinct aspects of the nonlinear response. Four ingredients enter the proposed picture.

\textit{(i) Cherenkov-like threshold.}
In a conventional drift picture, spontaneous acoustic emission becomes kinematically allowed when the carrier drift velocity exceeds the phase velocity of the relevant acoustic mode,
$v_{\mathrm{drift}}>s_{\mathrm{eff}}$,

where $v_{\mathrm{drift}}$ is the carrier drift velocity and $s_{\mathrm{eff}}$ is the effective phase velocity of the acoustic mode participating in the emission process.

In the hydrodynamic regime, electron--electron collisions establish local equilibrium, and the relevant quantity is the local velocity field of the electron fluid, $\mathbf{u}(\mathbf{r})$. The threshold condition then becomes local,
$|\mathbf{u}(\mathbf{r})|\gtrsim s_{\mathrm{eff}}$.

Both the kinematic condition $|\mathbf{u}|=s_{\mathrm{eff}}$ and the convective term $(\mathbf{u}\cdot\nabla)\mathbf{u}$ are even under current reversal. The convective term may characterize the magnitude and spatial localization of the nonlinear flow near the inner contact, but it cannot account for the observed polarity selectivity.

\textit{(ii) Corbino current concentration.}
For radial flow, the current density is
\begin{equation}
j_r(r)=\frac{I}{2\pi r},
\end{equation}

where $I$ is the total current and $r$ is the radial coordinate measured from the center of the Corbino disk.

The corresponding local hydrodynamic velocity is
\begin{equation}
u_r(r)=\frac{I}{2\pi rne},
\end{equation}

where $n$ is the two-dimensional carrier density and $e$ is the elementary charge.

Both the current density and the fluid velocity are largest at the inner radius, $r_{\mathrm{in}}$, so the supersonic condition is reached first at $r=r_{\mathrm{in}}=R_a=6~\mu\mathrm{m}$. Setting
$|u_r(r_{\mathrm{in}})|=s_{\mathrm{eff}}$
defines the local kinematic threshold current
\begin{equation}
I_{u=s}=2\pi r_{\mathrm{in}}nes_{\mathrm{eff}}.
\label{eq:Ius}
\end{equation}

For 

$n=n_s=7.1\times10^{11}\,\mathrm{cm}^{-2}$, where $n_s$ is the measured sheet carrier density,

and $s_{\mathrm{eff}}=3\text{--}5\,\mathrm{km/s}$,
Eq.~\eqref{eq:Ius} gives
$I_{u=s}\simeq130\text{--}215\,\mu\mathrm{A}$.

Equation~\eqref{eq:Ius} predicts a linear dependence of the local kinematic threshold on the inner-contact radius,
$I_{u=s}\propto r_{\mathrm{in}}$.
However, all Corbino devices studied here have the same inner radius,
$r_{\mathrm{in}}=6\,\mu\mathrm{m}$,
and differ only in their outer radius. The present measurements therefore probe the effect of the finite annulus width but do not provide a direct test of the predicted inner-radius dependence. A systematic device series with different $r_{\mathrm{in}}$ will be required to test this scaling and to determine whether a critical inner-contact diameter exists for observing a pronounced transport anomaly.

Equation~\eqref{eq:Ius} depends on the magnitude, but not on the sign, of the current. It therefore determines when acoustic-phonon emission becomes locally allowed, but does not explain why the corresponding differential-resistance feature is observed predominantly for one current polarity.

\textit{(iii) Corbino paradox and localized boundary drop.}
In an ideal radial Corbino flow, the bulk velocity profile is smooth, but the electrochemical-potential drop can be concentrated near the contacts~\cite{falkovich,kumar2}. Accordingly,
\begin{equation}
V(I)
=
V_{\mathrm{bulk}}(I)
+
\Delta V_{\mathrm{in}}(I)
+
\Delta V_{\mathrm{out}}(I),
\label{eq:Vdecomp}
\end{equation}

where $V(I)$ is the total measured voltage, $V_{\mathrm{bulk}}(I)$ is the voltage drop across the Corbino annulus, and $\Delta V_{\mathrm{in}}(I)$ and $\Delta V_{\mathrm{out}}(I)$ are the voltage contributions associated with the inner- and outer-contact regions, respectively.

The inner-contact contribution is expected to be particularly sensitive to a local threshold process because the normal current density is largest at $r_{\mathrm{in}}$. If a new dissipation channel associated with acoustic emission opens near the inner contact, its contribution may be written as
\begin{equation}
\Delta V_{\mathrm{in}}(I)
=
\Delta V_{\mathrm{in}}^{(0)}(I)
+
\Delta V_{\mathrm{Ch}}(I),
\end{equation}

where $\Delta V_{\mathrm{in}}^{(0)}(I)$ is the smooth inner-contact voltage contribution in the absence of the threshold-like anomaly and $\Delta V_{\mathrm{Ch}}(I)$ is the excess voltage associated with the Cherenkov-like acoustic-emission channel.

The corresponding excess differential-resistance contribution is
\begin{equation}
\Delta R_{\mathrm{peak}}(I)
\simeq
\frac{d\Delta V_{\mathrm{Ch}}(I)}{dI},
\label{eq:DRpeak}
\end{equation}

where $\Delta R_{\mathrm{peak}}(I)$ denotes the background-subtracted threshold-like contribution to the measured differential resistance.

\textit{(iv) Polarity selection by contact heating and cooling.}
A natural current-odd mechanism is reversible Peltier heating and cooling at the contacts. Upon current reversal, the inner contact changes from a locally heated boundary to a locally cooled boundary, or vice versa. Because the current density and the contact response are strongest at the inner radius, this thermoelectric contribution modifies the local electron temperature, energy-relaxation rate, and boundary conditions precisely in the region where the acoustic threshold is first reached.

This interpretation is supported by a parity-resolved analysis of the nonlinear magnetoresistance of the same Corbino devices, which reveals an approximately linear current-odd electron-temperature contribution consistent with the Peltier effect; a detailed thermal analysis will be reported separately. The acoustic-emission condition therefore remains kinematically available for both current directions, whereas the Peltier contribution determines the polarity for which the resulting excess dissipation becomes most visible.

\subsection{Quantitative analysis}

\subsubsection{Linear boundary scale}

For a current-penetrable boundary at zero field, the linear boundary condition is $U = j_n/G$, 
where $U$ is the electrochemical-potential jump across the boundary expressed in voltage units, $j_n$ is the normal current density,
and $G$ is the boundary conductance per unit length. Using $k_F = \sqrt{2\pi n_s} \simeq 2.1 \times 10^{8}\,\mathrm{m}^{-1}$ and $G \simeq 0.96 \times 2 G_S$, where $G_S = (2 e^2/h)(k_F/\pi)$ is the Sharvin conductance per unit length, the factor of 2 accounts for spin and the prefactor $0.96$ is the standard hydrodynamic correction~\cite{raichev}. This gives $G \simeq 1.0 \times 10^{4}\,\mathrm{S/m}$. The corresponding ideal inner-contact resistance is
\begin{equation}
R_{\mathrm{in}}^{\mathrm{Sh}}
=
\frac{1}{2\pi r_{\mathrm{in}}G}
\simeq 2.7\,\Omega,
\end{equation}
and the linear potential jump at $I \simeq 510\,\mu\mathrm{A}$ [$j_n(r_{\mathrm{in}}) \simeq 13.5\,\mathrm{A/m}$] is $U_{\mathrm{Sharvin}} \sim 1\text{--}2\,\mathrm{mV}$.

This sets the reference scale for an ideal linear current-penetrable boundary; it is two orders of magnitude smaller than the total measured contact resistance ($300\text{--}400\,\Omega$) and provides the linear-response baseline against which any nonlinear excess must be compared.

\subsubsection{Anomalous voltage scale}

The threshold-induced voltage is extracted by integrating the background-subtracted peak,
\begin{equation}
\Delta V_{\mathrm{Ch}}(I)
=
\int_{I_0}^{I}
\Delta R_{\mathrm{peak}}(I')\,dI',
\end{equation}
with $I_0$ taken outside the threshold region and
$\Delta R_{\mathrm{peak}}$ obtained after subtracting the third-order
phenomenological background described above (Fig.~4). Applied to the five Corbino datasets, this yields $\Delta V_{\mathrm{Ch}} \simeq 18\text{--}43\,\mathrm{mV}$, an order of magnitude larger than $U_{\mathrm{Sharvin}}$. The corresponding effective excess boundary resistance is $R_{\mathrm{Ch}}^{\mathrm{eff}} \sim \Delta V_{\mathrm{Ch}}/I_{\mathrm{peak}} \sim 35\text{--}85\,\Omega$, and the additional dissipated power is $P_{\mathrm{Ch}} \sim I_{\mathrm{peak}}\Delta V_{\mathrm{Ch}} \sim 9\text{--}22\,\mu\mathrm{W}$.

The extracted voltage is an \textit{electrochemical}-potential difference, $\mu_{\mathrm{ec}} = \mu(n) - e\varphi$, and not a purely electrostatic drop. Because Coulomb screening keeps the bulk 2DES close to charge neutrality, large density variations in the annulus are energetically unfavorable, and a substantial part of $\Delta\mu_{\mathrm{ec}}$ appears as a localized boundary jump at the current-penetrable inner contact. The large $\Delta V_{\mathrm{Ch}}$ therefore reflects a strongly nonlinear boundary condition rather than a static gate-like potential applied to the entire 2DES; the effective inner-contact jump may be written as
\begin{equation}
U_{\mathrm{eff}}(I)
=
U^{(0)}(I)
+
\Delta V_{\mathrm{Ch}}(I).
\end{equation}

\subsubsection{Visibility factor $\alpha_C$}

The observed peak occurs at $I_{\mathrm{peak}} \simeq 400\text{--}550\,\mu\mathrm{A}$, several times the bare kinematic threshold $I_{u=s}$. We summarize this offset by writing
\begin{equation}
I_{\mathrm{peak}}
=
\alpha_C I_{u=s}
=
\alpha_C\cdot 2\pi r_{\mathrm{in}}n e s_{\mathrm{eff}},
\label{eq:alphaCdef}
\end{equation}
with $\alpha_C \sim 2\text{--}4$.

The factor $\alpha_C$ should not be interpreted as a modification of the sound-barrier condition. Instead, it is a phenomenological transport-visibility factor relating the local kinematic onset to the experimentally resolved maximum in $R_{\mathrm{diff}}$. Immediately above $I_{u=s}$, the supersonic region has only a small radial extent and its contribution to the total differential resistance may remain too weak to resolve. A pronounced peak appears only when a sufficiently extended near-contact region becomes supersonic and the resulting acoustic dissipation becomes comparable to the smooth nonlinear background.

The geometric content is straightforward. For $|I|>I_{u=s}$, the radius
\begin{equation}
r_s(I)=\frac{|I|}{2\pi n e s_{\mathrm{eff}}}
\end{equation}
defines the outer boundary of the locally supersonic region. At the bare threshold this region has zero radial extent. For $|I|=\alpha_C I_{u=s}$, it would reach from $r_{\mathrm{in}}$ to $r_s=\alpha_C r_{\mathrm{in}}$, with area
\begin{equation}
A_{\mathrm{super}}
=
\pi r_{\mathrm{in}}^2(\alpha_C^2-1),
\end{equation}
provided this radius remains inside the outer contact. In the present devices $R_b/r_{\mathrm{in}}\approx1.8\text{--}3$, so for the smaller rings the supersonic region reaches the outer boundary and the relevant upper limit becomes
\begin{equation}
r_s=\min(\alpha_C r_{\mathrm{in}},R_b).
\end{equation}

Assuming a deformation-potential emission rate per unit area
\[
w_{\mathrm{em}}(u,s)
=
w_0\bigl(|u|-s\bigr)^2
\Theta\bigl(|u|-s\bigr),
\]

the acoustic power dissipated in the supersonic annulus is
\begin{equation}
P_{\mathrm{ac}}(I)
=
2\pi w_0 r_{\mathrm{in}}^2 s_{\mathrm{eff}}^2
f(x;\rho),
\qquad
x\equiv\frac{|I|}{I_{u=s}},
\quad
\rho\equiv\frac{R_b}{r_{\mathrm{in}}},
\label{eq:Pac}
\end{equation}
with
\begin{widetext}
\begin{equation}
f(x;\rho)
=
\int_{1}^{\min(x,\rho)}
\xi
\left(
\frac{x}{\xi}-1
\right)^2
d\xi
=
\begin{cases}
x^2\ln x-\tfrac{3}{2}x^2+2x-\tfrac{1}{2},
& x\leq\rho,\\[6pt]
x^2\ln\rho-2x(\rho-1)+\tfrac{1}{2}(\rho^2-1),
& x>\rho.
\end{cases}
\label{eq:fx}
\end{equation}
\end{widetext}

In the unbounded limit, $f(x)=x^2\ln x-\tfrac{3}{2}x^2+2x-\tfrac{1}{2}$ satisfies $f(1)=f'(1)=f''(1)=0$, so the channel switches on smoothly rather than as a jump at $x=1$. Once the supersonic region fills the annulus ($x>\rho$), $P_{\mathrm{ac}}$ grows more slowly, with $f(x;\rho)/x^2\rightarrow\ln\rho$ for $x\gg\rho$; this saturation may contribute to the non-monotonic dependence of $I_{\mathrm{peak}}$ on $R_b$.

Taking the peak to emerge once $P_{\mathrm{ac}}$ reaches a fraction $\eta$ of the background dissipation $P_{\mathrm{bg}}\simeq R_{\mathrm{bg}}I^2$ gives the implicit condition
\begin{equation}
\frac{f(\alpha_C;\rho)}{\alpha_C^2}
=
\eta
\frac{
R_{\mathrm{bg}}I_{u=s}^2
}{
2\pi w_0r_{\mathrm{in}}^2s_{\mathrm{eff}}^2
}
\equiv\Lambda.
\label{eq:visibility}
\end{equation}

The left-hand side rises monotonically from zero at $x=1$, with $f(x)/x^2\approx0.07,\ 0.21,\ 0.36$ at $x=2,\ 3,\ 4$ for an unbounded annulus, and somewhat smaller values once the finite-$R_b$ cutoff is included. The observed range $\alpha_C\sim2\text{--}4$ thus corresponds to a modest visibility ratio $\Lambda\sim0.07\text{--}0.35$: the acoustic channel becomes transport-visible when it contributes of order ten to thirty percent of the background dissipation, a plausible level for a peak to rise above a smooth polynomial background. This is only a phenomenological estimate within a simple visibility model---$w_0$, $\eta$, and $R_{\mathrm{bg}}$ are not known independently---and equation~\eqref{eq:visibility} should be read as a consistency check rather than a first-principles prediction.

A complementary, model-independent estimate follows from the power scales already extracted. The ratio of the anomalous voltage to the ideal Sharvin scale is $\Delta V_{\mathrm{Ch}}/U_{\mathrm{Sharvin}}\sim10\text{--}20$; since dissipation grows approximately as $I^2$, raising the relevant power scale by this factor requires a current larger by $\sqrt{10\text{--}20}\simeq3\text{--}4.5$, consistent with the observed $I_{\mathrm{peak}}/I_{u=s}\sim2.5\text{--}4$. Both estimates place $\alpha_C$ in the same range and support the interpretation that the offset between the kinematic and transport-visible thresholds reflects competition between acoustic emission and background dissipation, modulated by the finite Corbino width.

\subsection{Consistency with the experimental observations}

The polarity dependence must be distinguished from the acoustic-emission threshold itself. A homogeneous kinematic threshold based on $|u|=s_{\mathrm{eff}}$ is symmetric under current reversal, because changing the sign of $I$ reverses the direction of the velocity field without changing its magnitude. The wiring-reversal control described in the Supplemental Material demonstrates that the anomaly follows the physical current direction relative to the inner contact and is not caused by the polarity of the current source, voltage amplifier, or wiring configuration.

We attribute the remaining physical polarity selectivity to Peltier heating and cooling at the contacts. Reversing the current reverses the sign of the contact heat flux and therefore changes the local electron temperature and energy-relaxation conditions near the inner contact. The $1/r$ current concentration makes this thermoelectric modulation strongest precisely where the acoustic threshold is first reached. Thus, the Corbino geometry localizes the threshold process near the inner contact, whereas the current-odd Peltier response determines the polarity for which the acoustic-emission contribution becomes transport-visible. The present transport measurements support this contact-region interpretation, although they do not constitute direct spatial imaging of the emission region.

Application of a perpendicular magnetic field modifies the electron
trajectories, momentum and energy relaxation, and the local
electrochemical-potential distribution near the contacts. At the same
time, the large positive Corbino magnetoresistance increasingly dominates
$R_{\mathrm{diff}}$ near $I=0$ and progressively obscures and broadens
the threshold-like contribution. Both effects smooth and suppress the negative-current anomaly with increasing $|B|$. A residual polarity asymmetry nevertheless survives at finite $B$, indicating that the inner-contact inequivalence persists even as the bulk transport acquires increasingly classical-cyclotron character.

The Hall-bar control device confirms the geometric localization of the effect from the opposite direction. Its $R_{\mathrm{diff}}(I)$ is broad, smooth, and nearly symmetric under current reversal (Fig.~5), consistent with generic Joule heating or weak bulk electron--phonon relaxation but devoid of any sharp polarity-selective feature. The Hall bar lacks the $1/r$ current concentration required to produce a strongly localized supersonic region. The contrast between the two geometries shows that the Corbino threshold peak is not a generic high-current nonlinearity of the 2DES but is strongly enhanced by the radial current concentration and the associated contact-region physics.

The temperature dependence fits the same picture. As the bath temperature is raised, the threshold peak shifts to larger $|I|$ without significant narrowing. Within the present model, $T$ enters through the background dissipation $R_{\mathrm{bg}}$, which sets the scale that the acoustic channel must compete with, and through the local electron temperature and energy-relaxation conditions near the inner contact. Both can raise the visibility ratio $\Lambda$ in equation~\eqref{eq:visibility} and hence increase $\alpha_C$ together with the observed $I_{\mathrm{peak}}$. The acoustic velocities and carrier density vary negligibly over the studied range, so the kinematic threshold $I_{u=s}$ itself is essentially temperature independent; the observed temperature dependence resides predominantly in the transport-visibility factor.

\subsection{Ruling out alternative mechanisms}

The most natural alternative is straightforward Joule heating. Heating depends on $|I|^{2}$ and is therefore symmetric under current reversal, in clear conflict with the polarity selectivity of the Corbino peak and with the smooth, symmetric response observed in the Hall bar fabricated from the same wafer. Local heating undoubtedly contributes to the nonequilibrium state of the inner-contact region, but it cannot generate the sharp polarity-asymmetric onset on its own.

Impact ionization can be excluded on energy grounds. The relevant electrochemical-potential scale, $e\Delta V_{\mathrm{Ch}} \sim 20\text{--}40\,\mathrm{meV}$, is one to two orders of magnitude below any interband or intervalley ionization threshold in GaAs/AlGaAs at low temperature, and an ionization-driven onset would in any case be symmetric under current reversal.

A Gunn-like instability of the negative-differential-conductivity type is similarly excluded. GaAs has no NDC branch at the carrier densities and biases probed here: the $\Gamma$--$L$ intervalley transfer is reached only at electric fields of several kV/cm, whereas the local field in the Corbino at $I = I_{\mathrm{peak}}$, $E_r(r_{\mathrm{in}}) = \rho_0\, j_r(r_{\mathrm{in}}) \simeq 0.5\,\mathrm{V/cm}$, is four orders of magnitude smaller. Any such bulk instability would also not be confined to a single contact.

A purely contact-driven nonlinearity fails on several fronts at once: it cannot reproduce the threshold-like onset, the polarity-selective sharpness of the peak, the smooth magnetic-field suppression, or the systematic relation between $\Delta V_{\mathrm{Ch}}$ and the inner-contact current density. The contacts are demonstrably ohmic in the linear regime --- the low-bias $I$--$V$ characteristics are linear and the magnetoresistance is monotonic --- so the anomaly cannot be assigned to a generic nonlinear contact, even though it does originate in the strongly nonequilibrium boundary region adjacent to the inner contact.

Taken together, the data are consistent with a Corbino-assisted
Cherenkov-like threshold for acoustic phonon emission associated with the
inner-contact region and controlled by the transport-visibility
coefficient $\alpha_C$ defined by Eq.~(16).
\section{Conclusion}

We have studied nonlinear transport in GaAs/AlGaAs two-dimensional electron systems patterned in Corbino geometries and observe a pronounced polarity-selective threshold peak in the differential resistance, appearing predominantly on the negative-current branch and with no comparable sharp polarity-selective feature in the Hall-bar
device fabricated from the same wafer. 

We interpret the anomaly as a local, boundary-assisted Cherenkov-like threshold for acoustic phonon emission associated with the Corbino-enhanced local carrier velocity near the inner contact. The radial current concentration, $j_r(r)=I/(2\pi r)$, causes the local velocity to reach the sound velocity first in a region adjacent to the inner Corbino contact. 

Because both the condition $|u|=s$ and the convective term are even under current reversal, they cannot account for the observed polarity selectivity. We instead attribute the polarity selection to Peltier heating and cooling at the contacts, which modify the local electron temperature and boundary conditions near the inner contact.

The threshold position is captured by $I_{\mathrm{peak}}=\alpha_C I_{u=s}$ with a transport-visibility coefficient $\alpha_C\sim2$--$4$, which a simple phenomenological visibility model relates to the competition between the integrated acoustic dissipation in the radial flow and background nonlinear losses. 

Here, $I_{u=s}$ represents the local kinematic onset of an allowed emission channel, whereas $I_{\mathrm{peak}}$ corresponds to the experimentally resolved maximum that emerges only after a sufficiently extended supersonic region produces a detectable excess resistance.

The extracted threshold-induced electrochemical-potential drop, $\Delta V_{\mathrm{Ch}}\sim18$--$43\,\mathrm{mV}$, exceeds the ideal Sharvin boundary scale by more than an order of magnitude, consistent with a strongly nonlinear excess contribution associated with the inner-contact boundary region. 

A perpendicular magnetic field predominantly increases the low-bias Corbino resistance and progressively suppresses and broadens the narrow threshold-like peak, while a broader high-current asymmetry remains visible.

The absence of a comparable sharp feature in the Hall-bar control device and the non-monotonic dependence of $I_{\mathrm{peak}}$ on the outer radius support the importance of the radial current concentration and near-contact conditions rather than a spatially homogeneous bulk mechanism.

Beyond the specific Corbino response reported here, our results identify the Corbino geometry as a sensitive probe of local nonequilibrium electron--phonon processes in a high-mobility system operating within an independently established hydrodynamic-crossover regime. 

Hydrodynamics therefore provides the transport context and a natural framework for describing the local velocity field and current-penetrable boundary region, but the observed anomaly itself is not presented as proof of hydrodynamic transport.
The results suggest that the inner Corbino contact may act as a candidate geometry-defined source of supersonic acoustic emission. Combining such contacts with lithographically defined acoustic cavities or piezoelectric heterostructures may provide a route to compact, contact-defined acoustic phonon emitters in high-mobility two-dimensional electron systems.
\section {ACKNOWLEDGMENTS}
This work is supported by FAPESP (São Paulo Research Foundation) Grants No. 2019/16736-2, No. 2021/12470- 8, No. 2025/22786-3, CNPq (National Council for Scientific and Technological Development). The growth of GaAs quantum wells and preliminary transport measurements were supported by the Ministry of Science and Higher Education of the Russian Federation.
\section{DATA AVAILABILITY}
The data that support the findings of this article are openly
available \cite{zenodo}.


\clearpage
\onecolumngrid

\setcounter{equation}{0}
\setcounter{figure}{0}
\setcounter{table}{0}
\setcounter{page}{1}
\setcounter{section}{0}

\makeatletter
\renewcommand{\theequation}{S\arabic{equation}}
\renewcommand{\thefigure}{S\arabic{figure}}
\renewcommand{\thetable}{S\arabic{table}}
\renewcommand{\thepage}{S\arabic{page}}
\makeatother

\begin{center}
{\large\bfseries Supplemental Material for\\[0.5em]
Corbino-Enhanced Supersonic Acoustic-Emission Threshold in a GaAs
Two-Dimensional Electron System\par}

\vspace{1.0em}

{\normalsize
A. D. Levin$^{1}$, A. S. Jaroshevich$^{2}$, Z. D. Kvon$^{2,3}$,
V. A. Chitta$^{1}$, M. S. Aksenov$^{2}$, D. V. Dmitriev$^{2}$,
A. K. Bakarov$^{2}$, and G. M. Gusev$^{1}$\par}

\vspace{0.8em}

{\small
$^{1}$Instituto de F\'{\i}sica da Universidade de S\~ao Paulo,
135960-170, S\~ao Paulo, SP, Brazil\\
$^{2}$Institute of Semiconductor Physics, Novosibirsk 630090, Russia\\
$^{3}$Novosibirsk State University, Novosibirsk 630090, Russia\par}

\vspace{0.8em}
{\small \today\par}
\end{center}

\vspace{1.0em}
\noindent\textbf{Supplemental Material.}
The Supplemental Material presents additional measurements supporting the
reproducibility of the nonlinear Corbino response and the control experiments
discussed in the main text.

\vspace{1.0em}

The temperature evolution of the nonlinear response in these samples is
practically identical to that observed in the corresponding devices shown
in Fig.~2 of the main text. In particular, both additional samples exhibit
the same pronounced enhancement of $R_{\mathrm{diff}}$ on the
negative-current branch and a threshold-like feature at several hundred
microamperes. With increasing temperature, the position and overall shape
of the anomaly evolve in the same way as in the devices presented in the
main text. Small differences in the absolute resistance and peak position
can be attributed to sample-to-sample variations in the contact and
boundary conditions. These measurements demonstrate that the observed
temperature-dependent nonlinear response is reproducible in independent
Corbino devices.

Figure~\ref{fig:S2} presents a control measurement performed on the
$R_b=18~\mu\mathrm{m}$ Corbino device after physically interchanging both
the current and voltage connections between the inner and outer contacts.
The data are plotted using the same physical current direction relative to
the inner contact. The differential-resistance curves obtained before and
after interchanging the connections coincide within experimental accuracy.
In particular, the position, amplitude, and shape of the threshold-like
feature remain unchanged. The observed nonlinear response is therefore
independent of the particular wiring configuration and cannot be attributed
to the polarity of the current source, voltage amplifier, or other
instrumental asymmetry. Instead, the anomaly follows the physical direction
of current flow relative to the geometrically inequivalent inner contact.

\begin{figure}[H] 
\includegraphics[width=8cm]{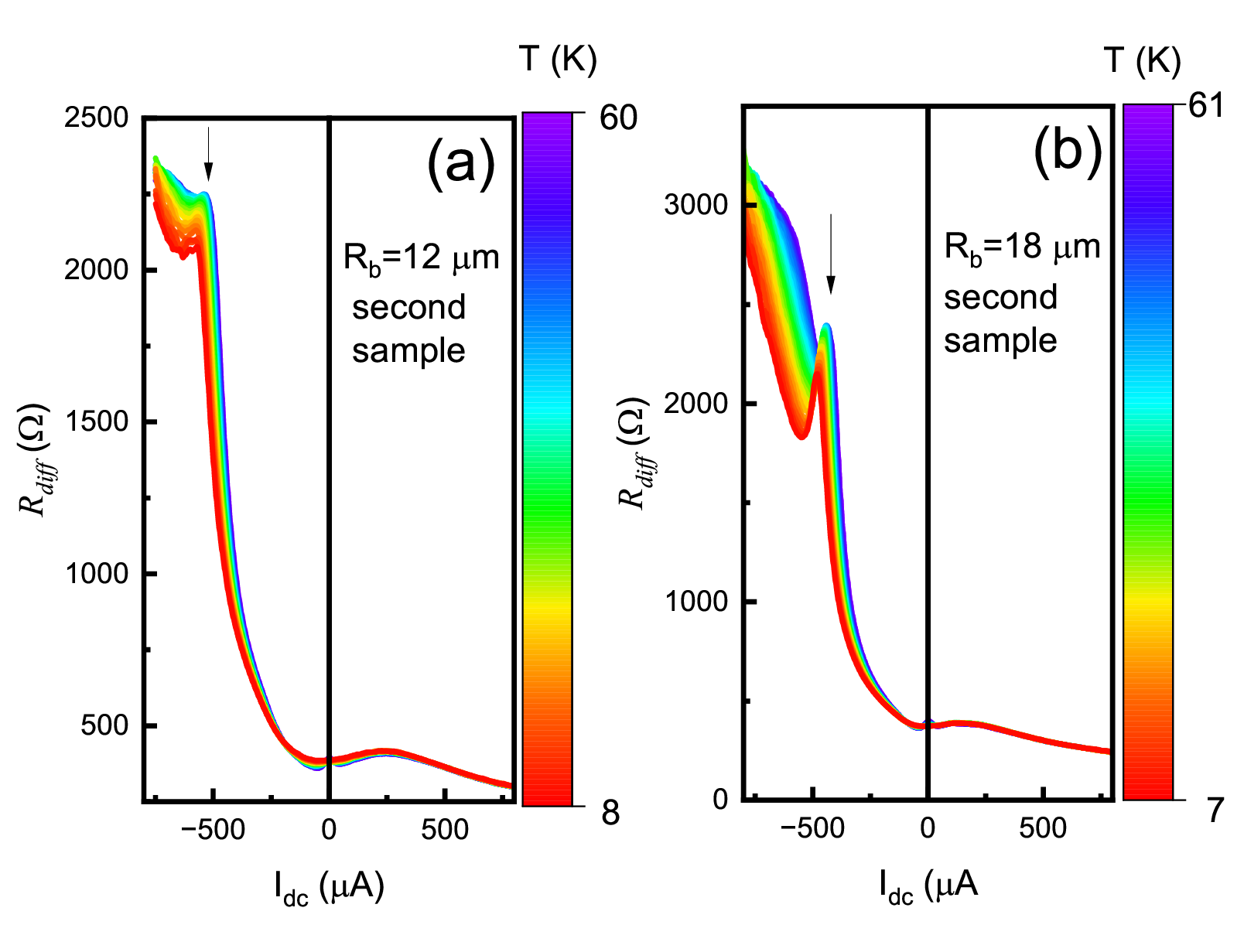}
\caption{(Color online)  Temperature evolution of the differential resistance
$R_{\mathrm{diff}}$ as a function of dc bias current for Corbino devices
with  $R_b=12~\mu\mathrm{m}$ (a), and
$R_b=18~\mu\mathrm{m}$ (b) for 2 additional samples. Each curve corresponds to a different bath temperature. Panels (a) and (b) contain 27, and 28 traces, respectively,
measured in steps of $2~\mathrm{K}$. Arrows indicate the positions of the peaks on the negative-current branch.}
 \label{fig:S1}
\end{figure}

\begin{figure}[t]
    \includegraphics[width=8 cm]{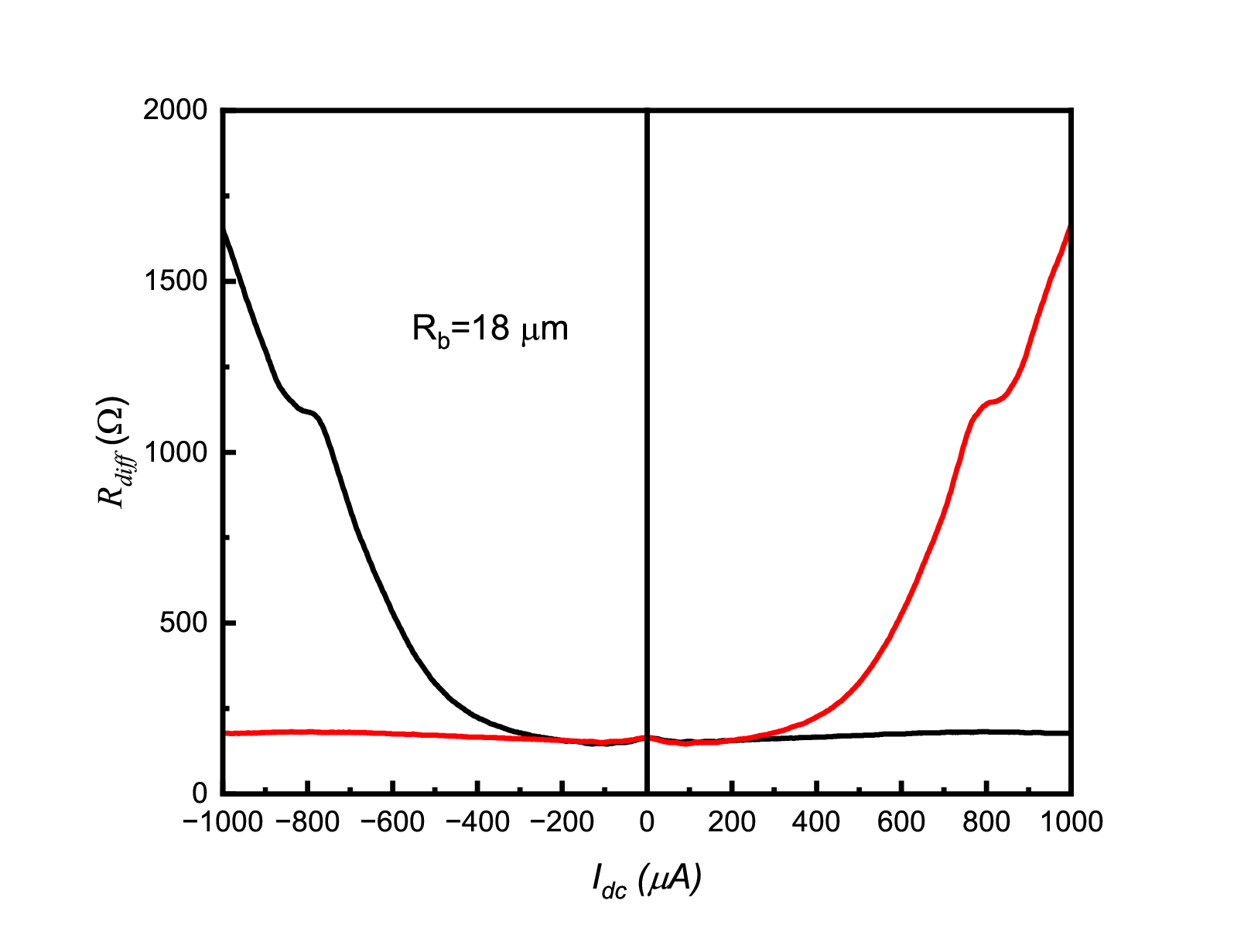}
    \caption{(Color online) Differential resistance
$R_{\mathrm{diff}}(I_{\mathrm{dc}})$ for the
$R_b=18~\mu\mathrm{m}$ Corbino device measured for two wiring
configurations. The black curve corresponds to the original wiring
configuration, for which the threshold-like anomaly appears on the
negative-current branch. The red curve corresponds to the configuration
obtained after physically interchanging both the current and voltage
connections between the inner and outer contacts. In the instrumental
current convention, the anomaly reverses its current polarity after the
connections are interchanged. Upon reversing the current sign for the
reversed configuration, the two measurements coincide within experimental
accuracy, demonstrating that the anomaly follows the physical direction
of current flow relative to the inner contact rather than the instrumental
polarity.}
    \label{fig:S2}
\end{figure}

\begin{figure}[t]
    \includegraphics[width=8 cm]{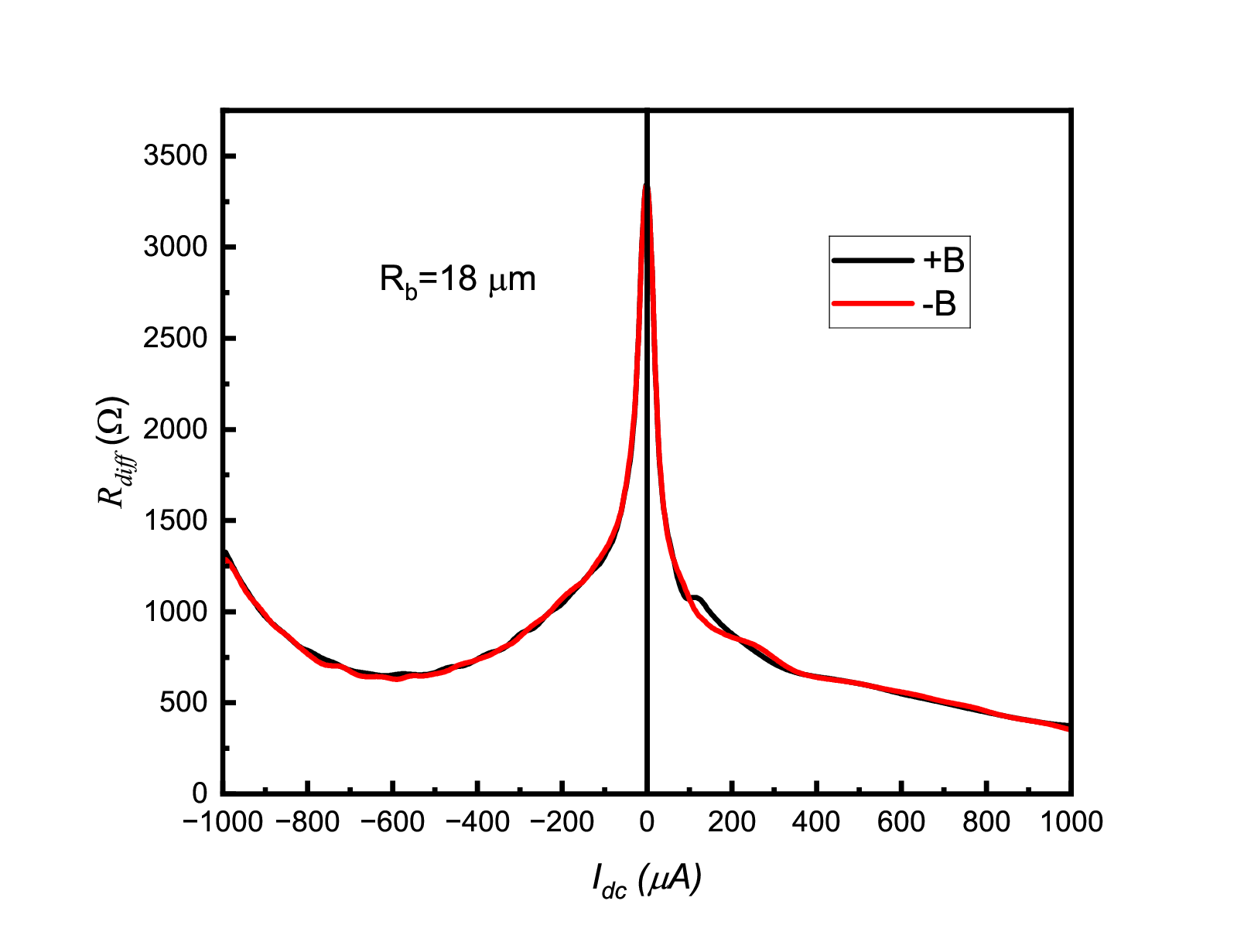}
    \caption{
    (Color online) Differential resistance
    $R_{\mathrm{diff}}(I_{\mathrm{dc}})$ for the
    $R_b=18~\mu\mathrm{m}$ Corbino device measured at
    $B=+0.5~\mathrm{T}$ and $B=-0.5~\mathrm{T}$.
    The two curves coincide within experimental accuracy, demonstrating
    that the nonlinear response is independent of the sign of the
    perpendicular magnetic field.
    }
    \label{fig:S3}
\end{figure}

Figure~\ref{fig:S3} shows the differential resistance of the
$R_b=18~\mu\mathrm{m}$ Corbino device measured at equal positive and
negative perpendicular magnetic fields,
$B=+0.5~\mathrm{T}$ and $B=-0.5~\mathrm{T}$.
The two $R_{\mathrm{diff}}(I_{\mathrm{dc}})$ curves coincide within
experimental accuracy over the measured current range. Both the
field-induced enhancement of the low-current resistance and the suppression
of the threshold-like feature are practically unchanged upon reversal of
the magnetic field. The nonlinear response therefore depends on the
magnitude of the perpendicular field rather than on its sign, within the
experimental accuracy. This control excludes an odd-in-$B$ contribution as
the origin of the observed threshold-like anomaly.

\end{document}